\documentclass{article}
\usepackage{spconf,amsmath,graphicx}

\usepackage{booktabs}
\usepackage{url}
\usepackage{cite}

\usepackage{amssymb}
\usepackage{mathtools, nccmath}

\usepackage{multirow}  

\title{FAST-RIR: FAST NEURAL DIFFUSE ROOM IMPULSE RESPONSE GENERATOR}
\name{Anton Ratnarajah$^1$, Shi-Xiong Zhang$^2$, Meng Yu$^2$, Zhenyu Tang$^1$, Dinesh Manocha$^1$, Dong Yu$^2$}
\address{$^1$ University of Maryland, College Park, MD, USA \\
         $^2$ Tencent AI Lab, Bellevue, WA, USA\\
         $^1$\{jeran, zhy, dmanocha\}@umd.edu, $^2$\{auszhang, raymondmyu, dyu\}@tencent.com}
\begin{document}
%
\maketitle
\begin{abstract}
We present a neural-network-based fast diffuse room impulse response generator (FAST-RIR) for generating room impulse responses (RIRs) for a given acoustic environment. Our FAST-RIR takes rectangular room dimensions, listener and speaker positions, and reverberation time ($T_{60}$) as inputs and generates specular and diffuse reflections for a given acoustic environment. Our FAST-RIR is capable of generating RIRs for a given input $T_{60}$ with an average error of 0.02s. We evaluate our generated RIRs in automatic speech recognition (ASR) applications using Google Speech API, Microsoft Speech API, and Kaldi tools. We show that our proposed FAST-RIR with batch size 1 is 400 times faster than a state-of-the-art diffuse acoustic simulator (DAS) on a CPU and gives similar performance to DAS in ASR experiments. Our FAST-RIR is 12 times faster than an existing GPU-based RIR generator (gpuRIR). We show that our FAST-RIR outperforms gpuRIR by 2.5\% in an AMI far-field ASR benchmark.

\end{abstract}
\section{Introduction}
\label{sec:intro}
 
Room impulse response (RIR) generators are used to simulate large-scale far-field speech training data \cite{google,convolv1,reverb_speech}. A synthetic far-field speech training dataset is created by convolving clean speech with RIRs generated for different acoustic environments and adding background noise \cite{convolv1,BUTREVERB}. The acoustic environment can be described using room geometry, speaker and listener positions, and room acoustic materials.

In recent years, an increasing number of RIR generators have been introduced to generate a realistic RIR for a given acoustic environment \cite{image_method,ray1,zhenyu_GAS,StoRIR}. Accurate RIR generators can generate RIRs with various acoustic effects (e.g., diffraction, scattering, early reflections, late reverberation) \cite{liu2020sound}. A limitation of accurate RIR generators is that they are computationally expensive, and the time taken to generate RIRs depends on the geometric complexity of the acoustic environment. Also, many ray-based RIR simulators use the empirical Sabine formula \cite{sabine} to compute the acoustic absorption coefficients from the desired reverberation time. Reverberation time ($T_{60}$) is the time required for the sound energy to decay by 60 decibels \cite{reverb_time}. 

With advancements in deep neural-network-based far-field speech processing, the demand for on-the-fly simulation of far-field speech training datasets with hundreds of thousands of room configurations similar to the testing environment is increasing \cite{augment1,gpurir,augment3,augment2}. The CPU-based offline simulation of far-field speech with balanced $T_{60}$ distribution requires a lot of computation time and disk space \cite{image_method,zhenyu_GAS}, thus it is not scalable for production-level ASR training. 
One strategy to improve the speed of RIR generation is parallelizing most of the stages in the existing RIR generators and making the algorithm compatible for running on GPUs \cite{gpurir,RIRusingGPU}. 


\noindent{\bf Main Contributions:} We propose a neural-network-based fast diffuse room impulse response generator (FAST-RIR) that can be directly controlled using rectangular room dimension, listener and speaker positions, and $T_{60}$. $T_{60}$ implicitly reflects the characteristics of the room materials such as the floor, ceiling, walls, furniture etc. Our FAST-RIR takes a constant amount of time to generate an RIR for any given acoustic environment, and yields accurate $T_{60}$. 


Our FAST-RIR architecture is trained to generate both specular and diffuse reflections for a given acoustic environment. Diffuse reflection is widely observed in real-world environments and it is important to accurately model RIR. We show that our FAST-RIR can generate RIRs 400 times faster than the state-of-the-art diffuse acoustic simulator (DAS) \cite{zhenyu_GAS} on a single CPU and 12 times faster than gpuRIR\cite{gpurir} on a single GPU. The RIRs generated using our FAST-RIR perform similarly to the RIRs generated using the DAS and outperform gpuRIR by up to 2.5\% in far-field automatic speech recognition (ASR) experiments. Our FAST-RIR can generate RIRs for a given input $T_{60}$ with an average error of 0.02s.

\section{Related Works}
\label{sec:related}
The RIR generators developed over the decades can be divided into three groups: wave-based, ray-based, and neural-network-based techniques. Wave-based techniques are designed to give the most accurate results by solving wave equations \cite{wave3,wave4}. However, the wave-based techniques are only feasible for generating RIRs for less complicated scenes at low frequencies. Ray-based techniques \cite{reb_savoja} are less accurate than the wave-based approach because the wave nature of the sound is neglected. The image method \cite{image_method} and diffuse acoustic simulators \cite{zhenyu_GAS} are commonly used ray-based methods in speech-related tasks. The image method only models specular reflections while diffuse acoustic simulators accurately model both diffuse and specular reflections. 

Recently, neural-network-based RIR generators \cite{ir-gan,tsrir,image2reverb} have been introduced to generate RIRs for a given acoustic environment. IR-GAN \cite{ir-gan} is a GAN-based RIR generator that is trained on a real-world RIR dataset to generate realistic RIRs. However, IR-GAN does not take conventional environmental parameters as input by design, making it less configurable than traditional RIR generators.
\begin{figure}[t] 
	\centering
	\includegraphics[width=\linewidth]{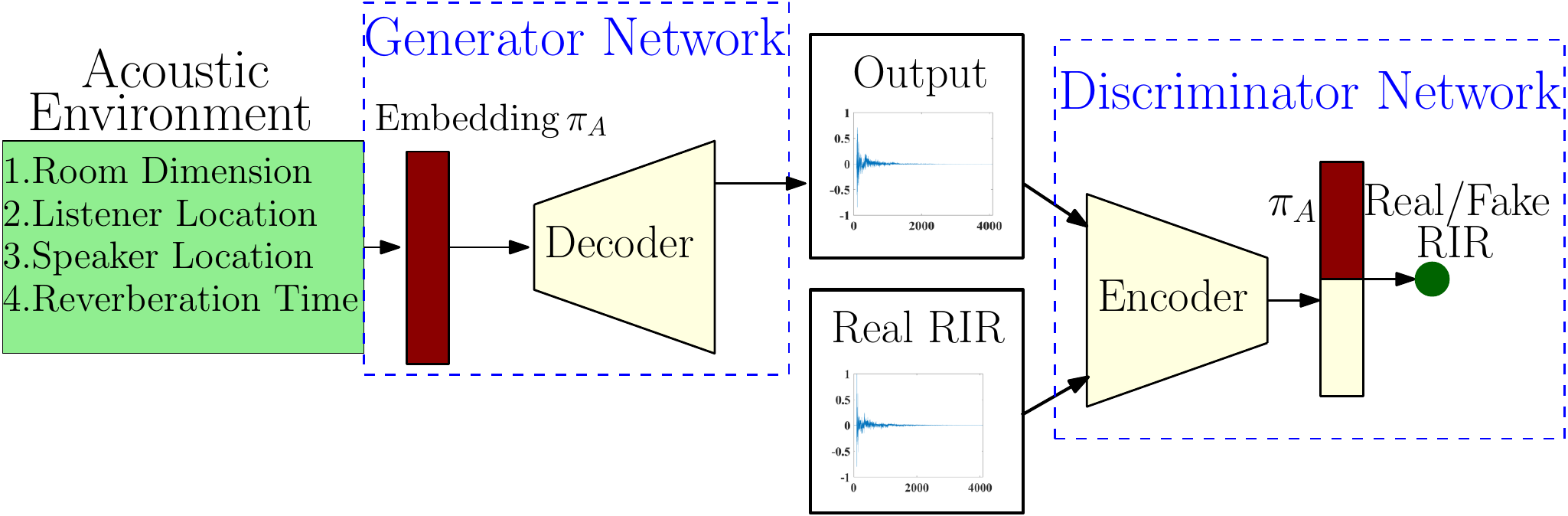}
	\caption{The architecture of our FAST-RIR. Our Generator network takes acoustic environment details as input and generates corresponding RIR as output. Our Discriminator network discriminates between the generated RIR and the ground truth RIR for the given acoustic environment during training.} 
	\label{architecture}
	\vspace{-0.4cm}
\end{figure}

\section{Our Approach}
\label{sec:approach}
To generate RIRs for a given acoustic environment, we propose a one-dimensional conditional generator network. Our generator network takes room geometry, listener and speaker positions, and $T_{60}$ as inputs, which are the common input used by all traditional RIR generators, and generates RIRs as raw-waveform audio. Our FAST-RIR generates RIRs of length 4096 at 16 kHz frequency.

\subsection{Modified Conditional GAN}
\label{sec:mod_GAN}
We propose a modified conditional GAN architecture to precisely generate an RIR for a given condition. GAN \cite{GAN} consists of a generator ($G$) and a discriminator ($D$) networks that are alternatingly trained to compete. The network $G$ is trained to learn a mapping from noise vector samples ($z$) from distribution $p_z$ to the data distribution $p_{data}$. The network $G$ is optimized to produce samples that are difficult for the $D$ to distinguish from real samples ($x$) taken from true data distribution, while $D$ is optimized to differentiate samples generated from $G$ and real samples. The networks $G$ and $D$ are trained to optimize the following two-player min-max game with value function $V(G,D)$.
{\belowdisplayskip 0.4\belowdisplayshortskip
\begin{equation}\label{gan_loss}
\begin{aligned}[b]
    \min_{G} \max_{D} V(G,D) = \mathbb{E}_{x \sim p_{data}} [\log D(x)] + \\
    \mathbb{E}_{z \sim p_{z}} [\log (1 - D(G(z)))].
\end{aligned}
\end{equation}
\vspace{-0.3cm}
}

Conditional  GAN (CGAN) \cite{conditional_GAN,conditional_gan2} is an extended version of GAN where both the generator and discriminator networks are conditioned on additional information $y$. The generator network in CGAN is conditioned on the random noise $z$ and $y$. The vector $z$ is used to generate multiple different samples satisfying the given condition $y$. In our work, we train our FAST-RIR to generate a single sample precisely for a given condition. Our FAST-RIR is a modified CGAN architecture where the generator network is only conditioned on $y$.

\subsection{FAST-RIR}
\label{sec:FAST-RIR}
We combine rectangular room dimension, listener location, and source location represented using 3D Cartesian coordinates ($x,y,z$) and $T_{60}$ as a ten-dimensional vector embedding $\pi_{A}$. We normalize the vector embedding within the range -1.2 to 1.2 using the largest room dimension in the training dataset.

For each $\pi_{A}$, we generate RIR using DAS ($R_{D}$) and use it as ground truth to train our network. Our objective function for the generator network ($G_{N}$) consists of modified CGAN error, mean square error and $T_{60}$ error. The discriminator network is trained using the modified CGAN objective function.

\subsubsection{Generator Modified CGAN Error}
The $G_{N}$ is trained with the following modified CGAN error to generate RIRs that are difficult for the discriminator $D_{N}$ to differentiate from RIRs generated from DAS.
{\belowdisplayskip 0.4\belowdisplayshortskip
\begin{equation}\label{CGAN_loss}
\begin{aligned}[b]
    \mathcal{L}_{CGAN} = \mathbb{E}_{\pi_{A} \sim p_{data}}[\log(1 - D_{N}(G_{N}(\pi_{A})))].
\end{aligned}
\end{equation}
\vspace{-0.5cm}
} 
\subsubsection{Mean Square Error (MSE)}
We compare each sample ($s$) of the RIR generated using our FAST-RIR ($R_{N}$) with RIR generated using DAS ($R_{D}$) for each $\pi_{A}$ to calculate the following MSE.
{\belowdisplayskip 0.4\belowdisplayshortskip
\begin{equation}\label{mse_loss}
\begin{aligned}[b]
    \mathcal{L}_{MSE} = \mathbb{E}_{\pi_{A} \sim p_{data}}[\mathbb{E}[(R_{N}(\pi_{A},s) - R_{D}(\pi_{A},s))^{2}]].
\end{aligned}
\end{equation}
\vspace{-0.5cm}
}
\subsubsection{$T_{60}$ Error}

We generate RIRs using our FAST-RIR and calculate their $T_{60}$ using a method based on ISO 3382-1:2009. We compare the $T_{60}$ of each generated RIRs with the $T_{60}$ given as input to the network in the embedding $\pi_{A}$ as follows:

{\belowdisplayskip 0.4\belowdisplayshortskip
\begin{equation}\label{t60_loss}
\begin{aligned}[b]
    \mathcal{L}_{T_{60}} = \mathbb{E}_{\pi_{A} \sim p_{data}}[|T_{60}(G_{N}(\pi_{A})) - T_{60}(\pi_{A})|].
\end{aligned}
\end{equation}
\vspace{-0.5cm}
}

\subsubsection{Full Objective}
We train the $G_{N}$ and $D_{N}$ alternatingly to minimize the generator objective function $\mathcal{L}_{G_{N}}$ (Equation \ref{generator_loss}) and maximize the discriminator objective function $\mathcal{L}_{D_{N}}$ (Equation \ref{discriminator_loss}). We control the relative importance of the MSE ($\mathcal{L}_{MSE}$) and $T_{60}$ error ($\mathcal{L}_{T_{60}}$) using the weights $\lambda_{MSE}$ and $\lambda_{T_{60}}$, respectively.

{\belowdisplayskip 0.4\belowdisplayshortskip
\begin{equation}\label{generator_loss}
\begin{aligned}[b]
    \mathcal{L}_{G_{N}} = \mathcal{L}_{CGAN} + \lambda_{MSE} \mathcal{L}_{MSE} + \lambda_{T_{60}} \mathcal{L}_{T_{60}}.
\end{aligned}
\end{equation}
} 
{\belowdisplayskip 0.4\belowdisplayshortskip
\begin{equation}\label{discriminator_loss}
\begin{aligned}[b]
    \mathcal{L}_{D_{N}} = \mathbb{E}_{(R_{D},\pi_{A}) \sim p_{data}}[\log(D_{N}(R_{D}(\pi_{A})))] \\
    + \mathbb{E}_{\pi_{A} \sim p_{data}}[\log(1 - D_{N}(G_{N}(\pi_{A})))].
\end{aligned}
\end{equation}
\vspace{-0.5cm}
} 

\subsubsection{Implementation}
\label{sec:implement}
\noindent\textbf{Network Architecture: } We adapt the generator network ($G_{N}$) and the discriminator network ($D_{N}$) proposed in Stage-I of StackGAN architecture \cite{stackgan} and modify the networks. StackGAN takes a text description and a noise vector as input and generates a photo-realistic two-dimensional (2D) image as output. Our FAST-RIR takes acoustic environment details as input and generates an RIR as a one-dimensional (1D) raw-waveform audio output. We flatten the 2D convolutions into 1D to process 1D RIR in both $G_{N}$ and $D_{N}$. 

Unlike photo-realistic images, raw-waveform audio exhibits periodicity. Donahue et al. \cite{WaveGAN} suggest that filters with larger receptive fields are needed to process low frequencies (large wavelength signals) in the audio. We improve the receptive field of the original $G_{N}$ and the encoder in $D_{N}$ by increasing the kernel size (i.e., 3$\times$3 2D convolution becomes length 41 1D convolution) and strides (i.e., stride 2$\times$2 becomes stride 4$\times$1). We also replace the upsampling layer and the following convolutional layer with a transposed convolutional layer.

\noindent\textbf{Dataset: }The sizes of the existing real-world RIR datasets \cite{Reverb,ACE,BUTREVERB} are insufficient to train our FAST-RIR. Therefore, we generate 75,000 medium-sized room impulse responses using a DAS \cite{zhenyu_GAS} to create a training dataset. We choose 15 evenly spaced room lengths within the range 8m to 11m, 10 evenly spaced room widths between 6m and 8m, and 5 evenly spaced room heights between 2.5m and 3.5m to generate RIRs. We position the speaker and the listener at random positions within the room and generate 100 different RIRs for each combination of room dimensions (15$\times$10$\times$5). The $T_{60}$ values of our training dataset are between 0.2s and 0.7s.

\noindent\textbf{Training: }We iteratively train $G_{N}$ and $D_{N}$ using RMSprop optimizer with batch size 128 and learning rate 8x$10^{-5}$. For every 40 epochs, we decay the learning rate by 0.7. 

\section{Experiment and Results}
\label{sec:Experiment}

\subsection{Baselines}
\label{sec:baseline}
We randomly select 30,000 different acoustic environments within the range of the training dataset (Section \ref{sec:implement}). We generate RIRs corresponding to the selected acoustic environments using image method \cite{image_method}, gpuRIR \cite{gpurir}, DAS \cite{zhenyu_GAS} and FAST-RIR to evaluate the performance of our proposed FAST-RIR. IR-GAN \cite{ir-gan} does not have the capability to precisely generate RIRs for a given speaker and listener positions; therefore, we did not use IR-GAN in our experiments. 

\subsection{Runtime}
\label{sec:runtime}
We evaluate the runtime for generating 30,000 RIRs using image method, gpuRIR, DAS and FAST-RIR on an Intel(R) Xenon(R) CPU E52699 v4 @ 2.20 GHz and a GeForce RTX 2080 Ti GPU (Table \ref{table1}). The gpuRIR is optimized to run on a GPU; therefore, we generate RIR using gpuRIR only on a GPU. For a fair comparison with CPU implementations of image-method and DAS, we also generate RIRs using our FAST-RIR with batch size 1 on a CPU.

From Table \ref{table1}, we can see that our proposed FAST-RIR with batch size 1 is 400 times faster than DAS \cite{zhenyu_GAS} on a CPU. Our FAST-RIR is optimized to run on a GPU. We compare the performance of our FAST-RIR with an existing GPU-based RIR generator gpuRIR \cite{gpurir}. We can see that gpuRIR performs better than our FAST-RIR with batch size 1, which is not the real use case of our generator. To our best knowledge, the gpuRIR does not leverage the batch parallelization while this was supported in our FAST-RIR. We can see that our proposed FAST-RIR with batch size 64 is 12 times faster than gpuRIR.

\begin{table}[t]
    \setlength{\tabcolsep}{1.5pt}
	\renewcommand{\arraystretch}{0.85} 
	\caption{The runtime for generating 30,000 RIRs using image method, gpuRIR, DAS, and our FAST-RIR. Our FAST-RIR significantly outperforms all other methods in runtime.}
	\label{table1}
	\centering
	\begin{tabular}{@{}lcccr@{}}	
		\toprule
		\textbf{RIR Generator} & \textbf{Hardware} &\textbf{Total Time}  & \textbf{Avg Time} \\
		\midrule
		DAS \cite{zhenyu_GAS}& CPU & 9.01x$10^{5}$s  & 30.05s\\
		Image Method \cite{image_method}& CPU & 4.49x$10^{3}$s  & 0.15s\\
		\textbf{FAST-RIR(Batch Size 1)} & \textbf{CPU} & \textbf{2.15x\boldmath$10^{3}$s}  & \textbf{0.07s}\\
		\midrule
		gpuRIR \cite{gpurir}& GPU & 16.63s   & 5.5x$10^{-4}$s\\
		FAST-RIR(Batch Size 1) & GPU & 34.12s & 1.1x$10^{-3}$s\\
		\textbf{FAST-RIR(Batch Size 64)} & \textbf{GPU} & \textbf{1.33s} & \textbf{4.4x\boldmath$10^{-5}$s}\\ 
		FAST-RIR(Batch Size 128) & GPU & 1.77s & 5.9x$10^{-5}$s\\ 
		\bottomrule
	\end{tabular}
	\vspace{-0.3cm}
\end{table}

\begin{table}[t]
	\renewcommand{\arraystretch}{0.8} 
	\caption{$T_{60}$ error of our FAST-RIR for 30,000 testing acoustic environments. We report the $T_{60}$ error for RIRs cropped at $T_{60}$ and full RIRs. We only crop RIRs with $T_{60}$ below 0.25s.} 
	\label{table_t60}
	\centering
	\begin{tabular}{@{}lcc@{}}	
		\toprule
		\textbf{\boldmath$T_{60}$ Range} & \textbf{Crop RIR at \boldmath$T_{60}$} & \textbf{\boldmath$T_{60}$ Error} \\
		\midrule
		0.2s - 0.25s & No & 0.068s \\
		\textbf{0.2s - 0.25s} & \textbf{Yes} & \textbf{0.033s} \\
		0.25s - 0.7s &  -   &0.021s \\
		\midrule
		0.2s - 0.7s &No &0.029s \\
		\textbf{0.2s - 0.7s} &\textbf{Yes} & \textbf{0.023s} \\
		\bottomrule
	\end{tabular}
	\vspace{-0.3cm}
\end{table}

\subsection{\boldmath$T_{60}$ Error}
\label{sec:t60}
Table \ref{table_t60} shows the $T_{60}$ error of the generated RIRs calculated using Equation \ref{t60_loss}. We can see that the testing $T_{60}$ error of our FAST-RIR is high for input $T_{60}$ below 0.25s (0.068s) when compared to the input $T_{60}$ greater than 0.25s (0.021s). 

Our FAST-RIR is trained to generate RIRs with durations slightly above 0.25s. For the input $T_{60}$ below 0.25s, the generated RIR has a noisy output between $T_{60}$ and 0.25s. We notice that cropping the generated RIRs at $T_{60}$ improves the overall $T_{60}$ error from 0.029s to 0.023s.


\subsection{Simulated Speech Comparison}
\label{sec:compare}
We simulate reverberant speech $x_r[t]$ by convolving clean speech $x_{c}[t]$ from the LibriSpeech test-clean dataset \cite{LibriSpeech} with different RIRs $r[t]$ (Equation \ref{reverberant_speech}).

{\belowdisplayskip 0.4\belowdisplayshortskip
\begin{equation}\label{reverberant_speech}
\begin{aligned}[b]
    x_{r}[t] = x_{c}[t] \circledast r[t].
\end{aligned}
\vspace{-0.5em}
\end{equation}
}

We decode the simulated reverberant speech using Google Speech API\footnote{\url{https://cloud.google.com/speech-to-text/}} and Microsoft Speech API\footnote{\url{https://azure.microsoft.com/en-us/services/cognitive-services/speech-services/}}. Table \ref{table_decode} shows the Word Error Rate (WER) of the decoded speech.
No text normalization was applied in both cases, as only the relative WER differences between different RIR generators are concerned. 
For Google Speech API, we report WER for the clean and reverberant LibriSpeech test sets that are successfully decoded. The results of each speech API show that compared with the reverberant speech simulated using traditional RIR generators, reverberant speech simulated using our FAST-RIR is closer to the reverberant speech simulated using DAS \cite{zhenyu_GAS}. We provide reverberant speech audio examples, spectrograms and the source code for reproducibility at github\footnote{ \url{https://anton-jeran.github.io/FRIR/}}.

\subsection{Far-field Automatic Speech Recognition}
\label{sec:far-field}
We want to ensure that our FAST-RIR generates RIRs that are better than or as good as existing RIR generators for ASR. We use the AMI corpus \cite{ami} for our far-field ASR experiments. AMI contains close-talk speech data recorded using Individual Headset Microphones (IHM) and distant speech data recorded using Single Distant Microphones (SDM). 

We use a modified Kaldi recipe~\footnote{\url{https://github.com/RoyJames/kaldi-reverb/}} to evaluate our FAST-RIR. The modified Kaldi recipe takes IHM data as the training set and tests the model using SDM data. The IHM data can be considered clean speech because the echo effects in IHM data are negligible when compared to SDM data. We augment far-field speech data by reverberating the IHM data with different RIR sets using Equation \ref{reverberant_speech}. The 30,000 RIRs generated using the image method, gpuRIR, DAS, and FAST-RIR are used in our experiment.

The IHM data consists of 687 long recordings. Instead of reverberating a speech recording using a single RIR, we do segment-level speech reverberation, as proposed in \cite{BUTREVERB}. We split each recording at the beginning of at least continuous 3 seconds of silence. We split at the beginning to avoid inter-segment reverberated speech overlapping. We can split IHM data into 17749 segments. We reverberate each segment using a randomly selected RIR from an RIR dataset (either image method, gpuRIR, DAS, DAS-cropped or our FAST-RIR).

Table \ref{table_ami} presents far-field ASR development and test WER for far-field SDM data. We can see that our FAST-RIR outperforms gpuRIR \cite{gpurir} by up to 2.5\% absolute WER. The DAS \cite{zhenyu_GAS} with full duration and the DAS cropped to have the same duration as our FAST-RIR (DAS-cropped) performs similarly in the far-field ASR experiment. We see that the performance of DAS and FAST-RIR has no significant difference. 

\begin{table}[t]
  \setlength{\tabcolsep}{1.8pt}
	\renewcommand{\arraystretch}{0.85} 

	\caption{Automatic speech recognition (ASR) results were obtained using Google Speech API and Microsoft Speech API. We simulate a reverberant speech testing dataset by convolving clean speech from the LibriSpeech dataset with different RIR datasets. We compare the reverberant speech simulated using the image method, gpuRIR and our FAST-RIR with the reverberant speech simulated using DAS. We show that the relative WER change from our method is the smallest.}
	\label{table_decode}
	\centering
	\begin{tabular}{@{}llccc@{}}	
		\toprule
		\multicolumn{2}{l}{{\textbf{Testing Dataset}}}
			& \multicolumn{2}{c}{\textbf{Word Error Rate} [\%]}\\
\cmidrule(r{1pt}){3-4} 
		&\textbf{Clean Speech $\circledast$ RIR}	&\textbf{Google API} &\textbf{Microsoft API}\\  
		\midrule
		&Libri $\circledast$ DAS (baseline) \cite{zhenyu_GAS}& 6.56 & 2.63 \\
		\midrule
		&Libri $\circledast$ gpuRIR \cite{gpurir}& 9.39 (+43\%) & 3.78 (+44\%) \\
		&Libri $\circledast$ Image Method \cite{image_method}& 9.03 (+38\%) &  3.86 (+47\%)\\
		&\textbf{Libri} $\circledast$ \textbf{FAST-RIR (ours)} & \textbf{7.14 (+9\%)} & \textbf{2.76 (+5\%)} \\
		\bottomrule
	\end{tabular}
	\vspace{-0.2cm}
\end{table} 

\begin{table}[t]
  \setlength{\tabcolsep}{1.8pt}
	\renewcommand{\arraystretch}{0.85} 
	\caption{Far-field ASR results were obtained for far-field speech data recorded by single distance microphones (SDM) in the AMI corpus. The best results are shown in \textbf{bold}.}
	\label{table_ami}
	\centering
	\begin{tabular}{@{}llcc@{}}	
		\toprule
		\multicolumn{2}{l}{{\textbf{Training Dataset}}}
			& \multicolumn{2}{c}{\textbf{Word Error Rate} [\%]}\\
\cmidrule(r{1pt}){3-4} 
		&\textbf{Clean Speech $\circledast$ RIR}	&\textbf{dev} &\textbf{eval}\\  
		\midrule
		&IHM $\circledast$ None & 55.0 & 64.2 \\
		&IHM $\circledast$ Image Method \cite{image_method} & 51.7 & 56.1 \\
		&IHM $\circledast$ gpuRIR \cite{gpurir} & 52.2 & 55.5 \\
		\midrule
		&IHM $\circledast$ DAS \cite{zhenyu_GAS} & 47.9 & \textbf{52.5} \\
		&IHM $\circledast$ DAS-cropped \cite{zhenyu_GAS} & 48.3 & \textbf{52.6} \\
		&\textbf{IHM \boldmath$\circledast$ FAST-RIR (ours)} & 47.8 & \textbf{53.0} \\
			
		\bottomrule
	\end{tabular}
	\vspace{-0.3cm}
\end{table} 

\section{Discussion and Future Work}
\label{sec:discussion}
We propose a novel FAST-RIR architecture to generate a large RIR dataset on the fly. 
We show that our FAST-RIR performs similarly in ASR experiments when compared to the RIR generator (DAS \cite{zhenyu_GAS}), which is used to generate a training dataset to train our FAST-RIR. 
Our FAST-RIR can be easily trained with RIR generated using any state-of-the-art accurate RIR generator to improve its performance in ASR experiments while keeping the speed of RIR generation the same.

Although we trained our FAST-RIR for limited room dimensions ranging from (8m,6m,2.5m) to (11m,8m,3.5m) using 75,000 RIRs, we believe that our FAST-RIR will give a similar performance when we train FAST-RIR for a larger room dimension range with a huge amount of RIRs. We would like to evaluate the performance of our FAST-RIR in the multi-channel ASR \cite{multi-asr} and speech separation \cite{multi-seperate} tasks.

\bibliographystyle{IEEEbib.bst}
\bibliography{strings,refs}

\begin{thebibliography}{10}

\bibitem{google}
Chanwoo Kim, Ananya Misra, Kean Chin, Thad Hughes, Arun Narayanan, Tara~N.
  Sainath, and Michiel Bacchiani,
\newblock ``{Generation of Large-Scale Simulated Utterances in Virtual Rooms to
  Train Deep-Neural Networks for Far-Field Speech Recognition in Google
  Home},''
\newblock in {\em Proc. Interspeech 2017}, 2017, pp. 379--383.

\bibitem{convolv1}
Tom Ko, Vijayaditya Peddinti, Daniel Povey, Michael~L. Seltzer, and Sanjeev
  Khudanpur,
\newblock ``A study on data augmentation of reverberant speech for robust
  speech recognition,''
\newblock in {\em {ICASSP}}. 2017, pp. 5220--5224, {IEEE}.

\bibitem{reverb_speech}
Tara N.~Sainath et~al.,
\newblock ``Multichannel signal processing with deep neural networks for
  automatic speech recognition,''
\newblock {\em {IEEE} {ACM} Trans. Audio Speech Lang. Process.}, vol. 25, no.
  5, pp. 965--979, 2017.

\bibitem{BUTREVERB}
Igor Sz{\"{o}}ke, Miroslav Sk{\'{a}}cel, Ladislav Mosner, Jakub Paliesek, and
  Jan~Honza Cernock{\'{y}},
\newblock ``Building and evaluation of a real room impulse response dataset,''
\newblock {\em {IEEE} J. Sel. Top. Signal Process.}, vol. 13, no. 4, pp.
  863--876, 2019.

\bibitem{image_method}
Jont~B. {Allen} and David~A. {Berkley},
\newblock ``{Image method for efficiently simulating small-room acoustics},''
\newblock {\em Acoustical Society of America Journal}, vol. 65, no. 4, pp.
  943--950, Apr. 1979.

\bibitem{ray1}
Carl Schissler and Dinesh Manocha,
\newblock ``Interactive sound propagation and rendering for large multi-source
  scenes,''
\newblock {\em ACM Trans. Graph.}, vol. 36, no. 1, Sept. 2016.

\bibitem{zhenyu_GAS}
Zhenyu Tang, Lianwu Chen, Bo~Wu, Dong Yu, and Dinesh Manocha,
\newblock ``Improving reverberant speech training using diffuse acoustic
  simulation,''
\newblock in {\em {ICASSP}}. 2020, pp. 6969--6973, {IEEE}.

\bibitem{StoRIR}
Piotr Masztalski, Mateusz Matuszewski, Karol Piaskowski, and Michal Romaniuk,
\newblock ``Storir: Stochastic room impulse response generation for audio data
  augmentation,''
\newblock in {\em {INTERSPEECH}}. 2020, pp. 2857--2861, {ISCA}.

\bibitem{liu2020sound}
Shiguang Liu and Dinesh Manocha,
\newblock ``Sound synthesis, propagation, and rendering: a survey,''
\newblock {\em arXiv preprint arXiv:2011.05538}, 2020.

\bibitem{sabine}
Wallace~Clement Sabine and M~David Egan,
\newblock ``Collected papers on acoustics,'' 1994.

\bibitem{reverb_time}
Heinrich Kuttruff,
\newblock {\em Room acoustics},
\newblock Spon Press, 2009.

\bibitem{augment1}
Martin Karafi{\'{a}}t, Karel Vesel{\'{y}}, Katerina Zmol{\'{\i}}kov{\'{a}},
  Marc Delcroix, Shinji Watanabe, Luk{\'{a}}s Burget, Jan~Honza Cernock{\'{y}},
  and Igor Sz{\"{o}}ke,
\newblock ``Training data augmentation and data selection,''
\newblock in {\em New Era for Robust Speech Recognition, Exploiting Deep
  Learning}, pp. 245--260. Springer, 2017.

\bibitem{gpurir}
David Diaz-Guerra, Antonio Miguel, and Jose~R Beltran,
\newblock ``gpurir: A python library for room impulse response simulation with
  gpu acceleration,''
\newblock {\em Multimedia Tools and Applications}, vol. 80, no. 4, pp.
  5653--5671, 2021.

\bibitem{augment3}
Zhenyu Tang and Dinesh Manocha,
\newblock ``Scene-aware far-field automatic speech recognition,'' 2021.

\bibitem{augment2}
Martin Karafiát, František Grézl, Lukáš Burget, Igor Szöke, and Jan
  Černocký,
\newblock ``{Three ways to adapt a CTS recognizer to unseen reverberated speech
  in BUT system for the ASpIRE challenge},''
\newblock in {\em Proc. Interspeech 2015}, 2015, pp. 2454--2458.

\bibitem{RIRusingGPU}
Zhong{-}Hua Fu and Jian{-}Wei Li,
\newblock ``Gpu-based image method for room impulse response calculation,''
\newblock {\em Multim. Tools Appl.}, vol. 75, no. 9, pp. 5205--5221, 2016.

\bibitem{wave3}
D.~Botteldooren,
\newblock ``Acoustical finite‐difference time‐domain simulation in a
  quasi‐cartesian grid,''
\newblock {\em The Journal of the Acoustical Society of America}, vol. 95, no.
  5, pp. 2313--2319, 1994.

\bibitem{wave4}
D.~Botteldooren,
\newblock ``Finite‐difference time‐domain simulation of low‐frequency
  room acoustic problems,''
\newblock {\em The Journal of the Acoustical Society of America}, vol. 98, no.
  6, pp. 3302--3308, 1995.

\bibitem{reb_savoja}
Lauri Savioja and U.~Peter Svensson,
\newblock ``Overview of geometrical room acoustic modeling techniques,''
\newblock {\em The Journal of the Acoustical Society of America}, vol. 138, no.
  2, pp. 708--730, 2015.

\bibitem{ir-gan}
Anton Ratnarajah, Zhenyu Tang, and Dinesh Manocha,
\newblock ``{IR-GAN: Room Impulse Response Generator for Far-Field Speech
  Recognition},''
\newblock in {\em Proc. Interspeech 2021}, 2021, pp. 286--290.

\bibitem{tsrir}
Anton Ratnarajah, Zhenyu Tang, and Dinesh Manocha,
\newblock ``Ts-rir: Translated synthetic room impulse responses for speech
  augmentation,''
\newblock {\em arXiv preprint arXiv:2103.16804}, 2021.

\bibitem{image2reverb}
Nikhil Singh, Jeff Mentch, Jerry Ng, Matthew Beveridge, and Iddo Drori,
\newblock ``Image2reverb: Cross-model reverb impulse response synthesis,''
\newblock in {\em ICCV}, October 2021.

\bibitem{GAN}
Ian~J. Goodfellow, Jean Pouget{-}Abadie, Mehdi Mirza, Bing Xu, David
  Warde{-}Farley, Sherjil Ozair, Aaron~C. Courville, and Yoshua Bengio,
\newblock ``Generative adversarial nets,''
\newblock in {\em {NIPS}}, 2014, pp. 2672--2680.

\bibitem{conditional_GAN}
Mehdi Mirza and Simon Osindero,
\newblock ``Conditional generative adversarial nets,''
\newblock {\em arXiv preprint arXiv:1411.1784}, 2014.

\bibitem{conditional_gan2}
Jon Gauthier,
\newblock ``Conditional generative adversarial networks for convolutional face
  generation,''
\newblock in {\em Tech Report}, 2015.

\bibitem{stackgan}
Han Zhang, Tao Xu, Hongsheng Li, Shaoting Zhang, Xiaogang Wang, Xiaolei Huang,
  and Dimitris Metaxas,
\newblock ``Stackgan: Text to photo-realistic image synthesis with stacked
  generative adversarial networks,''
\newblock in {\em {ICCV}}, 2017.

\bibitem{WaveGAN}
Jesse~H. Engel, Kumar~Krishna Agrawal, Shuo Chen, Ishaan Gulrajani, Chris
  Donahue, and Adam Roberts,
\newblock ``Gansynth: Adversarial neural audio synthesis,''
\newblock in {\em {ICLR} (Poster)}. 2019, OpenReview.net.

\bibitem{Reverb}
Keisuke~Kinoshita et~al.,
\newblock ``The {REVERB} challenge: {A} benchmark task for reverberation-robust
  {ASR} techniques,''
\newblock in {\em New Era for Robust Speech Recognition, Exploiting Deep
  Learning}, pp. 345--354. Springer, 2017.

\bibitem{ACE}
James Eaton, Nikolay~D. Gaubitch, Alastair~H. Moore, and Patrick~A. Naylor,
\newblock ``The {ACE} challenge - corpus description and performance
  evaluation,''
\newblock in {\em {WASPAA}}. 2015, pp. 1--5, {IEEE}.

\bibitem{LibriSpeech}
Vassil Panayotov, Guoguo Chen, Daniel Povey, and Sanjeev Khudanpur,
\newblock ``Librispeech: An {ASR} corpus based on public domain audio books,''
\newblock in {\em {ICASSP}}. 2015, pp. 5206--5210, {IEEE}.

\bibitem{ami}
J~Carletta~et al.,
\newblock ``The ami meeting corpus: A pre-announcement,''
\newblock in {\em Proceedings of the Second International Conference on Machine
  Learning for Multimodal Interaction}. 2005, MLMI'05, p. 28–39,
  Springer-Verlag.

\bibitem{multi-asr}
A.S.~Subramanian et~al.,
\newblock ``Directional {ASR:} {A} new paradigm for {E2E} multi-speaker speech
  recognition with source localization,''
\newblock in {\em {ICASSP}}. 2021, pp. 8433--8437, {IEEE}.

\bibitem{multi-seperate}
Rongzhi Gu, Shi{-}Xiong Zhang, Yuexian Zou, and Dong Yu,
\newblock ``Complex neural spatial filter: Enhancing multi-channel target
  speech separation in complex domain,''
\newblock {\em {IEEE} Signal Process. Lett.}, vol. 28, pp. 1370--1374, 2021.

\end{thebibliography}

\end{document}